\documentclass[letterpaper,11pt]{article}%
\usepackage{amsmath}
\usepackage{epsfig}
\usepackage{graphicx}%
\usepackage{amsfonts}%
\usepackage{amssymb}

\begin{document}

\title{{\huge Very low energy matching of effective meson theories with QCD}\\\ \\\ \ \\\ }
\author{Christopher Smith
\and \ \ \ \ \ \ \ \medskip\\\textsl{Institut de Physique Th\'{e}orique, Universit\'{e} Catholique de
Louvain,}\linebreak \\\ \ \textsl{\linebreak Chemin du Cyclotron, 2, B-1348, Louvain-la-Neuve, Belgium}\\\ \ }
\date{July 22, 2002}
\maketitle

\begin{abstract}
A simple matching procedure is proposed to extract constraints on effective
meson theories. In this way, a QCD prediction for the pion decay constant is
found, $F_{\pi}=2m_{\pi}/\pi\approx90$ MeV. The same procedure also determines
other mesonic observables, like the decay width of the sigma meson to two photons.

Finally, some information which can be gained about the hadronic
light-by-light contributions to the muon anomalous moment are briefly commented.

\pagebreak 

\end{abstract}

When the electron field is integrated out from the QED action, one gets an
effective theory where only photons can propagate. In addition, virtual
fermion loops generate an infinite tower of self-interactions among the
photons. Schwinger calculated the one-loop effective action, for
\textit{constant} electromagnetic fields \cite{Schwinger}, to all orders%
\begin{equation}
\mathcal{L}_{eff,fermion}^{(1)}=-\dfrac{1}{8\pi^{2}}%
{\displaystyle\int_{0}^{\infty}}
\dfrac{d\tau}{\tau^{3}}e^{-m^{2}\tau}\left[  \left(  e\tau\right)
^{2}ab\dfrac{\cosh\left(  ea\tau\right)  \cos\left(  eb\tau\right)  }%
{\sinh\left(  ea\tau\right)  \sin\left(  eb\tau\right)  }-1\right]
\label{SchwingerFermion}%
\end{equation}
with $a,b$ solutions of $a^{2}-b^{2}=\mathbf{E}^{2}-\mathbf{B}^{2}$ and
$ab=\mathbf{E}\cdot\mathbf{B}$. When expanded with respect to the fermion mass
(or $\alpha$), one gets the four-photon interactions as described by the
well-known Euler-Heisenberg effective Lagrangian \cite{Euler}
\[
\mathcal{L}_{EH,fermion}^{(1)}=\frac{\alpha^{2}}{90m^{4}}\left(  \left(
F_{\mu\nu}F^{\mu\nu}\right)  ^{2}+\frac{7}{4}\left(  F_{\mu\nu}\tilde{F}%
^{\mu\nu}\right)  ^{2}\right)
\]
with the definition $\widetilde{F}^{\mu\nu}=\frac{1}{2}\varepsilon^{\mu\nu
\rho\sigma}F_{\rho\sigma}$. Diagrammatically, the integration of the fermion
fields can be represented as%
\[%
{\includegraphics[
height=1.2557in,
width=3.5042in
]%
{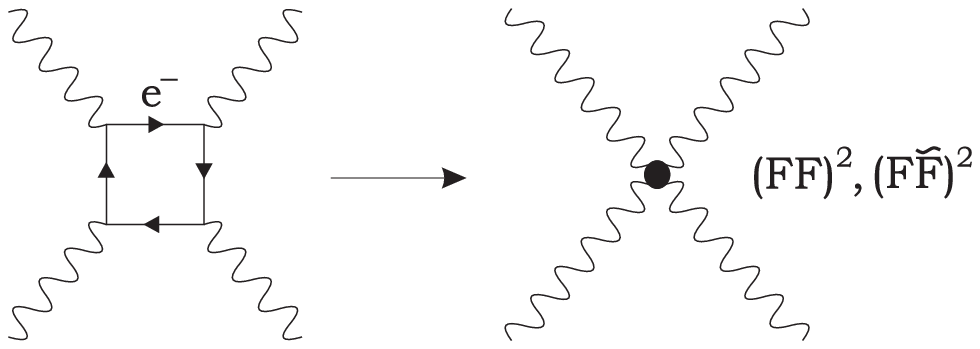}%
}%
\]
The same treatment can be applied to get the photon effective couplings
generated by virtual quark loops
\begin{equation}
\mathcal{L}_{EH,quarks}^{(1)}=e_{Q}^{4}\frac{\alpha^{2}N_{c}}{90m_{Q}^{4}%
}\left(  \left(  F_{\mu\nu}F^{\mu\nu}\right)  ^{2}+\frac{7}{4}\left(
F_{\mu\nu}\widetilde{F}^{\mu\nu}\right)  ^{2}\right)  \label{QuarkEH}%
\end{equation}
where $N_{c}$ is the number of colors, $e_{Q}$ is the quark charge (in unit of
$e$) and $m_{Q}$ its mass.

The starting point of the present letter is to assume that the relative
strength of the $\left(  F_{\mu\nu}F^{\mu\nu}\right)  ^{2}$ and $(F_{\mu\nu
}\widetilde{F}^{\mu\nu})^{2}$ couplings is preserved through the complicate
dressing by QCD of the quarks into hadrons. In other words, the hypothesis is
that strong corrections will be absorbed into the quark mass, the only free
parameter. The consequence is then that the same effective interaction among
the photons should be obtainable starting from an effective meson theory. By
matching the photon effective theories obtained by integrating out meson
fields to that obtained by integrating out the quark fields (\ref{QuarkEH}),
we will get a set of constraints on the parameters of the effective meson
theory. Ultimately, the validity of the basic hypothesis will be tested by
comparison with experiment.

At very low energy, a few $eV$ say, the contribution from pions will dominate.
We will use the scalar QED Lagrangian for the charged pions%
\[
\mathcal{L}_{\pi^{\pm}}=\partial_{\mu}\pi^{+}\partial^{\mu}\pi^{-}-m_{\pi}%
^{2}\pi^{+}\pi^{-}-ieA_{\mu}\left(  \pi^{+}\partial_{\mu}\pi^{-}-\partial
_{\mu}\pi^{+}\pi^{-}\right)  +e^{2}A_{\mu}A^{\mu}\pi^{+}\pi^{-}%
\]
and for the neutral pions, we introduce the coupling to two photons%
\[
\mathcal{L}_{\pi^{0}}=\frac{1}{2}\left[  \partial_{\mu}\pi^{0}\partial^{\mu
}\pi^{0}-m_{\pi}^{2}\pi^{0}\pi^{0}\right]  +g_{\pi}F_{\mu\nu}\widetilde
{F}^{\mu\nu}\pi^{0}%
\]
We now integrate out the pion fields from $\mathcal{L}_{\pi^{\pm}}%
+\mathcal{L}_{\pi^{0}}$. This generates contact interactions among photons:%
\[%
{\includegraphics[
height=2.3687in,
width=3.4627in
]%
{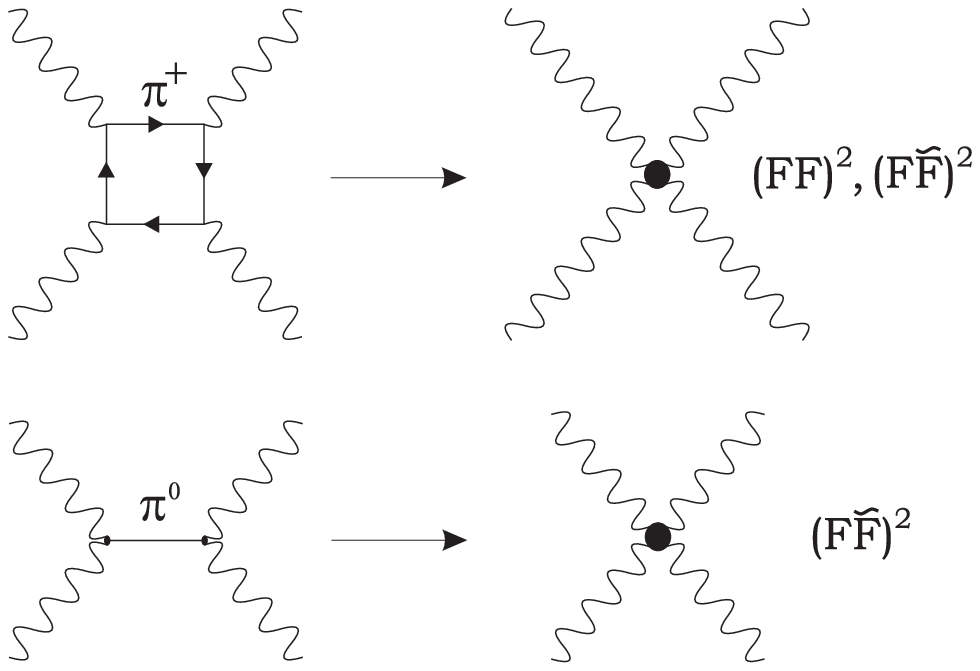}%
}%
\]
(it is understood that seagull interaction contributions are included).

Schwinger computed the effective action obtained by integrating out the
charged pions, with the result \cite{Schwinger}%
\begin{equation}
\mathcal{L}_{eff,\pi^{\pm}}^{(1)}=\dfrac{1}{16\pi^{2}}%
{\displaystyle\int_{0}^{\infty}}
\dfrac{d\tau}{\tau^{3}}e^{-m^{2}\tau}\left[  \left(  e\tau\right)
^{2}ab\dfrac{1}{\sinh\left(  ea\tau\right)  \sin\left(  eb\tau\right)
}-1\right]  \label{SchwingerScalar}%
\end{equation}
When expanded, the four-photon effective couplings are generated%
\begin{equation}
\mathcal{L}_{EH,\pi^{\pm}}^{(1)}=\frac{\alpha^{2}}{90m_{\pi^{\pm}}^{4}}\left(
\frac{7}{16}\left(  F_{\mu\nu}F^{\mu\nu}\right)  ^{2}+\frac{1}{16}\left(
F_{\mu\nu}\tilde{F}^{\mu\nu}\right)  ^{2}\right)  \label{ChargedPion}%
\end{equation}
To this effective Lagrangian, we add the effective interaction generated by a
neutral pion exchange%
\begin{equation}
\mathcal{L}_{EH,\pi^{0}}^{(1)}=\frac{g_{\pi}^{2}}{2m_{\pi^{0}}^{2}}\left(
F^{\mu\nu}\widetilde{F}_{\mu\nu}\right)  ^{2} \label{NeutralPion}%
\end{equation}
Combining (\ref{ChargedPion}) with (\ref{NeutralPion}), we get%
\begin{equation}
\mathcal{L}_{EH,\pi}^{(1)}=\frac{7}{16}\frac{\alpha^{2}}{90m_{\pi^{\pm}}^{4}%
}\left(  F_{\mu\nu}F^{\mu\nu}\right)  ^{2}+\left(  \frac{g_{\pi}^{2}}%
{2m_{\pi^{0}}^{2}}+\frac{1}{16}\frac{\alpha^{2}}{90m_{\pi^{\pm}}^{4}}\right)
\left(  F_{\mu\nu}\tilde{F}^{\mu\nu}\right)  ^{2} \label{PionEH}%
\end{equation}

The core of the method is to match this effective Lagrangian to the fermionic
one (\ref{QuarkEH}), to extract a prediction for $g_{\pi}$%
\[
\frac{\frac{g_{\pi}^{2}}{2m_{\pi^{0}}^{2}}+\frac{1}{16}\frac{\alpha^{2}%
}{90m_{\pi^{\pm}}^{4}}}{\frac{7}{16}\frac{\alpha^{2}}{90m_{\pi^{\pm}}^{4}}%
}=\frac{7}{4}\rightarrow g_{\pi}=\frac{\alpha}{8}\frac{m_{\pi^{0}}}%
{m_{\pi^{\pm}}^{2}}%
\]
For the decay rate $\pi^{0}\rightarrow\gamma\gamma$, this gives%
\[
\Gamma\left(  \pi^{0}\rightarrow\gamma\gamma\right)  =\frac{m_{\pi^{0}}^{3}%
}{4\pi}g_{\pi}^{2}=\frac{\alpha^{2}}{256\pi}\frac{m_{\pi^{0}}^{5}}{m_{\pi
^{\pm}}^{4}}=7.7\;eV
\]
To be compared to the experimental value $\Gamma^{\exp}\left(  \pi
^{0}\rightarrow\gamma\gamma\right)  =\left(  7.7\pm0.6\right)  $ eV
\cite{PDG}. The agreement is very good.

If the description of $\pi^{0}\rightarrow\gamma\gamma$ in terms of the axial
anomaly is used (\cite{Adler}, \cite{Bell}), we get a determination of
$F_{\pi}$%
\begin{equation}
g_{\pi}=\frac{\alpha N_{c}}{12\pi F_{\pi}}=\frac{\alpha}{8}\frac{m_{\pi^{0}}%
}{m_{\pi^{\pm}}^{2}}\rightarrow\fbox{$F_{\pi}=\dfrac{2N_{c}}{3\pi}%
\dfrac{m_{\pi^{\pm}}^{2}}{m_{\pi^{0}}}=91.9$ MeV} \label{Fpi}%
\end{equation}
again very close to the experimental value $F_{\pi}=92.4\pm0.3$ MeV (obtained
from $\Gamma\left(  \pi^{+}\rightarrow\mu^{+}\nu_{\mu}\right)  $, see
\cite{PDG}).

A comment is in order. Usually, in chiral perturbation theory, the role played
by $m_{\pi}$ and $F_{\pi}$ is radically different: $F_{\pi}$ sets the scale of
the Goldstone boson interactions, while $m_{\pi}$ is only a small explicit
breaking of the spontaneously broken symmetry; their physical content is
therefore quite different and a priori unrelated. On the other hand, if the
fermionic character of the underlying theory is assumed to be preserved
through the passage from the quark picture to the hadron picture, the two turn
out to be proportional.

As a by-product, we can also estimate the value of the constituent quark mass
for which the two descriptions match, i.e. for which the four-photon couplings
are the same in absolute magnitude. Setting $m_{u}=m_{d}$%
\[
e_{u}^{4}\frac{\alpha^{2}N_{c}}{90m_{u}^{4}}+e_{d}^{4}\frac{\alpha^{2}N_{c}%
}{90m_{d}^{4}}=\frac{7}{16}\frac{\alpha^{2}}{90m_{\pi^{\pm}}^{4}}\rightarrow
m_{u}=m_{\pi^{\pm}}\sqrt[4]{\left(  e_{u}^{4}+e_{d}^{4}\right)  \frac{48}{7}%
}\approx1.1\times m_{\pi^{\pm}}%
\]
i.e. a constituent quark mass $m_{u}=m_{d}\approx153\;$MeV, quite close to
$m_{\pi}$.\newline \newline 

\begin{center}
{\LARGE Introducing Higher Mass Particles\medskip\medskip\medskip\\[0pt]}
\end{center}

In principle, one can introduce the $\eta,\eta^{\prime},K^{\pm},...$, to get
the effective Lagrangian%
\begin{align}
\mathcal{L}_{EH,PS}^{(1)} &  =\left(  \frac{g_{\pi}^{2}}{2m_{\pi^{0}}^{2}%
}+\frac{g_{\eta}^{2}}{2m_{\eta}^{2}}+\frac{g_{\eta^{\prime}}^{2}}%
{2m_{\eta^{\prime}}^{2}}+\frac{\alpha^{2}}{1440}\left(  \frac{1}{m_{\pi^{\pm}%
}^{4}}+\frac{1}{m_{K^{\pm}}^{4}}\right)  \right)  \left(  F_{\mu\nu}\tilde
{F}^{\mu\nu}\right)  ^{2}\nonumber\\
&  +\frac{7\alpha^{2}}{1440}\left(  \frac{1}{m_{\pi^{\pm}}^{4}}+\frac{1}%
{m_{K^{\pm}}^{4}}\right)  \left(  F_{\mu\nu}F^{\mu\nu}\right)  ^{2}%
\label{AllPs}%
\end{align}
with $g_{\eta^{\left(  \prime\right)  }}$ the coupling constant for
$\eta^{\left(  \prime\right)  }\rightarrow\gamma\gamma$. We see that the pion
contributions are by far the dominant ones: the corrections induced by heavier
mesons are of a few percents, as can be seen by plugging in the experimental
values $g_{\eta}\approx6\times10^{-6}\,MeV^{-1}$ and $g_{\eta^{\prime}}%
\approx8\times10^{-6}\,MeV^{-1}$ \cite{PDG}. Matching (\ref{AllPs}) to
(\ref{QuarkEH}) and solving for $F_{\pi}$, we find $F_{\pi}\approx96$ MeV.
This shows that the matching should not be expected to work to a better
accuracy than roughly 5\%, despite the striking result (\ref{Fpi}).

We now turn to the introduction of resonances, and in particular of the sigma
meson. Our point of view is to consider the sigma as a resonance occurring in
the two-pion channel. We take as effective Lagrangian%
\begin{equation}
\mathcal{L}_{EH,\pi^{0},\sigma}^{(1)}=\frac{g_{\sigma}^{2}}{2m_{\sigma}^{2}%
}\left(  F_{\mu\nu}F^{\mu\nu}\right)  ^{2}+\frac{g_{\pi}^{2}}{2m_{\pi^{0}}%
^{2}}\left(  F_{\mu\nu}\tilde{F}^{\mu\nu}\right)  ^{2} \label{SigmaEH}%
\end{equation}
with $g_{\sigma}$ the coupling constant for $\sigma\rightarrow\gamma\gamma$.
To get this form, we assume that in a first approximation, the scalar channel
(i.e. $\left(  F_{\mu\nu}F^{\mu\nu}\right)  ^{2}$) is saturated by the sigma,
and we neglect the charged pion contribution to the pseudoscalar channel
$(F_{\mu\nu}\tilde{F}^{\mu\nu})^{2}$ (which is roughly 10 times smaller than
the neutral pion contribution, see (\ref{PionEH})). The effective photon
Lagrangian (\ref{SigmaEH}) is now only generated by virtual $\pi^{0}$ and
$\sigma$ exchanges, which are particles respectively associated with the axial
\cite{Adler} and trace anomaly \cite{Chanowitz}.

By matching (\ref{SigmaEH}) with (\ref{QuarkEH}), we can relate $g_{\sigma}$
to $g_{\pi}$%
\begin{equation}
\frac{g_{\sigma}^{2}/2m_{\sigma}^{2}}{g_{\pi}^{2}/2m_{\pi^{0}}^{2}}%
=\frac{4}{7}\rightarrow\frac{g_{\sigma}}{g_{\pi}}=\sqrt{\frac{4}{7}%
}\frac{m_{\sigma}}{m_{\pi^{0}}}\approx2.8 \label{gsigma}%
\end{equation}
for $m_{\sigma}\approx530$ MeV \cite{Ishida}. From (\ref{gsigma}), the sigma
width to two photons is predicted to be%
\begin{equation}
\Gamma\left(  \sigma\rightarrow\gamma\gamma\right)  =\frac{m_{\sigma}^{3}%
}{4\pi}g_{\sigma}^{2}=\frac{1}{7\pi}\frac{m_{\sigma}^{5}}{m_{\pi^{0}}^{2}%
}g_{\pi}^{2}=\frac{\alpha^{2}}{448\pi}\frac{m_{\sigma}^{5}}{m_{\pi^{\pm}}^{4}%
}\approx4.1_{-2.1}^{+3.5}\text{ keV} \label{SigmaWidth}%
\end{equation}
with $m_{\sigma}=\left(  530\pm70\right)  $ MeV \cite{Ishida}. This is
compatible with
\begin{equation}
\Gamma\left(  \sigma\rightarrow\gamma\gamma\right)  =\left(  3.8\pm1.5\right)
\,\text{keV } \label{BogliPenn}%
\end{equation}
found by a partial-wave analysis of $\gamma\gamma\rightarrow\pi\pi$
\cite{BoglionePennington}. Note, by the way, that because $m_{\sigma}$ appears
to the fifth power in (\ref{SigmaWidth}), this formula is in fact quite
efficient in constraining $m_{\sigma}$ once $\Gamma\left(  \sigma
\rightarrow\gamma\gamma\right)  $ is known. For instance, from the
experimental value (\ref{BogliPenn}), one finds $m_{\sigma}=520_{-49}^{+36}$ MeV.

In conclusion, the assumption that the sigma resonance nearly saturates the
two-pion scalar channel gives a reasonable prediction for $\Gamma\left(
\sigma\rightarrow\gamma\gamma\right)  $. \newline \newline 

\begin{center}
{\LARGE Conclusions and Perspectives\medskip\medskip\medskip\\[0pt]}
\end{center}

In this letter, we have shown how to get information on the coupling constant
of effective meson theories from QCD. As said, the present framework relies
entirely on the assumption that the strong interactions do not renormalize the
relative strength of the $\left(  F_{\mu\nu}F^{\mu\nu}\right)  ^{2}$ and
$(F_{\mu\nu}\tilde{F}^{\mu\nu})^{2}$ couplings (or at least that this
renormalization is small). The fact that we found reasonable predictions for
$F_{\pi}$ and $\Gamma\left(  \sigma\rightarrow\gamma\gamma\right)  $ seems to
validate this assumption. Further work should clarify the range of validity of
the present method.

Encouraged by the success of the matching at $O\left(  \alpha^{2}%
/m^{4}\right)  $, one could now undertake an analysis at the next order. At
$O\left(  \alpha^{3}/m^{8}\right)  $, the fermionic effective Lagrangian is
\cite{Schwinger}, \cite{DKR}
\[
\mathcal{L}_{DKR,quarks}^{(1)}=e_{Q}^{6}N_{c}\frac{\pi\alpha^{3}}{315m_{Q}%
^{8}}\left(  -4\left(  F_{\mu\nu}F^{\mu\nu}\right)  ^{3}-\frac{13}{2}\left(
F_{\alpha\beta}F^{\alpha\beta}\right)  \left(  F_{\mu\nu}\widetilde{F}^{\mu
\nu}\right)  ^{2}\right)
\]
At that order, the integration of the pseudoscalars is more involved, and it
remains to be seen which kind of constraints can emerge. Also, the extension
to non-constant electromagnetic fields may lead to interesting results. All
these questions are left for future studies.

We close this letter with a comment on the computation of the hadronic
light-by-light corrections to the muon anomalous moment. Even if it is true
that the momentum configuration in that case and in our case is quite
different, our approach can offer an interesting limiting case in which the
various theoretical models can be tested. For instance, in the present
approach, it appears that the pions and the constituent quarks do not
contribute simultaneously. The same is true for the sigma meson and the
charged pions. In the context of the muon anomalous moment, the constituent
quark, charged pion and sigma meson contributions are all considered at the
same time (see for example \cite{Kinoshita}). Whether this leads to
double-counting or not is, in our opinion, not settled.

{\large Acknowledgments}: I am grateful to J. Pestieau and S. Trine for useful
discussions. I acknowledge financial support from the I.I.S.N.


\begin{thebibliography}{9}                                                                                                %

\bibitem {Schwinger}J. Schwinger, Phys. Rev. \textbf{82}, 664 (1951).

\bibitem {Euler}H. Euler, W. Heisenberg, Z. Phys. \textbf{98}, 714 (1936).

\bibitem {PDG}D.E. Groom \textit{et al}, Eur. Phys. J. \textbf{C15}, 1 (2000).

\bibitem {Adler}S. L. Adler, Phys. Rev. \textbf{177}, 2426 (1969).

\bibitem {Bell}J.S. Bell, R. Jackiw, Nuovo Cim. \textbf{60A}, 46 (1969).

\bibitem {Chanowitz}M. S. Chanowitz, J. Ellis, Phys. Rev. \textbf{D7}, 2490 (1973).

\bibitem {Ishida}M. Ishida, S. Ishida, T. Komada and S.-I. Matsumoto,
proceedings of 9th International Conference on Hadron Spectroscopy (Hadron
2001), Protvino, Russia (\textit{hep-ph/0110358}), and references cited there.

\bibitem {BoglionePennington}M. Boglione, M.R. Pennington, Eur. Phys. J.
\textbf{C9}, 11 (1999).

\bibitem {DKR}D. Dicus, C. Kao, W. Repko, Phys. Rev. \textbf{D57}, 2443 (1998)

\bibitem {Kinoshita}E. de Rafael, Phys. Lett. \textbf{B322}, 239 (1994); M.
Hayakawa, T. Kinoshita, A. I. Sanda, Phys. Rev. \textbf{D54}, 3137 (1996).
\end{thebibliography}
\end{document}